\documentclass[english,prl, twocolumn]{revtex4-1}
\usepackage[T1]{fontenc}
\usepackage[latin9]{inputenc}
\usepackage{amsmath}
\usepackage{amssymb}
\usepackage{graphicx}

\makeatletter
\makeatletter
\renewcommand*{\fnum@figure}{{\normalfont\bfseries \figurename~\thefigure}}
\makeatother

\makeatother

\usepackage{babel}
\begin{document}
\title{Integrable Field Theories with an Interacting Massless Sector}
\author{Michael C. Abbott}
\affiliation{Holographic QFT Group, Wigner Research Centre for Physics, ~\\
Konkoly-Thege Miklós út. 29-33, 1121 Budapest, Hungary.}
\email{michael.abbott@wigner.mta.hu}

\author{In\^es Aniceto}
\affiliation{Mathematical Sciences, University of Southampton, ~\\
Highfield, Southampton SO17 1BJ, England.\vspace{3mm}}
\email{I.Aniceto@soton.ac.uk}

\date{27 February 2020}
\begin{abstract}
\noindent We present the first known integrable relativistic field
theories with interacting massive and massless sectors. And we demonstrate
that knowledge of the massless sector is essential for understanding
of the spectrum of the massive sector. Terms in this spectrum polynomial
in the spatial volume (the accuracy for which the Bethe ansatz would
suffice in a massive theory) require not just Lüscher-like corrections
(usually exponentially small) but the full TBA integral equations.
We are motivated by the implications of these ideas for AdS/CFT, but
present here only field-theory results.
\end{abstract}
\maketitle
%% NOTATION
%\newcommand{\thetak}{\theta_{(k)}} % 
%\newcommand{\thetaone}{\theta_{(1)}}
%\newcommand{\ennk}{n_{(k)}}
%\newcommand{\ennone}{n_{(1)}}
%\newcommand{\peek}{p_{(k)}}
%\newcommand{\peeone}{p_{(1)}}

\newcommand{\thetak}{\hat{\theta}_{k}}
\newcommand{\thetaj}{\hat{\theta}_{j}}
\newcommand{\thetaone}{\hat{\theta}_{1}}
\newcommand{\ennk}{\hat{n}_{k}}
\newcommand{\ennone}{\hat{n}_{k}}
\newcommand{\peek}{\hat{p}_{k}}
\newcommand{\peej}{\hat{p}_{j}}
\newcommand{\peeone}{\hat{p}_{k}}

\subsection*{Introduction}

\noindent Integrable quantum field theories are an important class
of exactly solvable models. Many, like the $O(N)$ and sine-Gordon
models \citep{Hasenfratz:1990hq,Zamolodchikov:1995xk}, contain massive
excitations, whose asymptotic S-matrix is the basic ingredient for
their solution. Some contain instead massless excitations, whose S-matrix
plays the same role \citep{Zamolodchikov:1991vx}. Among models which
contain both, or have adjustable mass parameters, it is generally
believed that the massless sector decouples, and so the two sectors
can be studied independently.

This paper studies some theories for which this decoupling does not
occur. We find a double-scaling limit of certain Homogeneous Sine-Gordon
(HSG) models \citep{Park:1994bx,Hollowood:1994vx,FernandezPousa:1996hi},
in which some of the particles become truly massless, yet retain a
nontrivial interaction with the massive particles. And we show that
the massless virtual particles must be included in the calculation
of the spectrum of massive excitations. We discuss examples in which
the full Thermodynamic Bethe Ansatz (TBA) \citep{Yang:1966ty,Zamolodchikov:1989cf}
is required, and one for which Lüscher terms \citep{Luscher:1985dn}
are sufficient. 

Our main motivation for seeking out such theories comes from the AdS/CFT
correspondence in string theory \citep{Maldacena:1997re}. In the
planar limit, this is a complicated integrable model, and techniques
of integrability have enabled the calculation of various quantities
far beyond either weak- or strong-coupling perturbation theory \citep{Beisert:2010jr}.
One version of this correspondence involves strings on $AdS_{3}\times S^{3}\times T^{4}$,
where the presence of the flat torus $T^{4}$ introduces massless
excitations into the (light-cone gauge) string theory \citep{Babichenko:2009dk,Borsato:2014hja}.
Fully incorporating these into the integrable description is the principal
challenge of adapting what we know about the $AdS_{5}\times S^{5}$
correspondence to this less-symmetric variant. The difficulties of
doing so have left open various disagreements concerning the energy
of massive physical states \citep{Abbott:2015pps}, and our hope is
that this papers's simpler examples may shed some light. 

\subsection*{$\mathbf{su(3)_{2}}$ Model}

\noindent Homogeneous Sine-Gordon (HSG) models are an integrable
family generalising the complex sine-gordon model \citep{Park:1994bx,Hollowood:1994vx,FernandezPousa:1996hi,Miramontes:1999hx,CastroAlvaredo:1999em,Dorey:2004qc,Bajnok:2015eng}.
We consider $su(3)_{k}$ models, all of which have three adjustable
parameters: $m_{i}$ for $i=1,2$ control the masses of the particles,
and $\sigma$ is a rapidity offset. The simplest model for which our
double-scaling limit exists is the $su(3)_{2}$ model. Its S-matrix
is diagonal, with $S^{ij}$ as follows:
\[
S(\theta)=\begin{bmatrix}-1 & \tanh\frac{1}{2}(\theta+\sigma-\frac{\mathrm{i}\pi}{2})\\
-\tanh\frac{1}{2}(\theta-\sigma-\frac{\mathrm{i}\pi}{2}) & -1
\end{bmatrix}.
\]
The vacuum TBA of this model was studied by \citep{CastroAlvaredo:1999em},
and following \citep{Dorey:1996re} we extend this to obtain excited-state
TBA equations with physical particles of mass $m_{1}$ \footnote{Our notation is that $\theta$ is a generic rapidity, and $\thetak$
that of a physical excitation, $\hat{E}_{k}=\sqrt{m_{1}^{2}+\peek^{2}}=m_{1}\cosh\thetak$.
We write $\phi$ specifically for the rapidity of particles of mass
$m_{2}$, as we will take the limit $m_{2}\to0$.}:
\begin{align}
\epsilon_{1}(\theta) & =m_{1}L\cosh\theta+\sum_{k}\pi\mathrm{i}-\int\frac{d\phi}{2\pi}\:K^{12}(\theta-\phi)L_{2}(\phi)\nonumber \\
\epsilon_{2}(\phi) & =m_{2}L\cosh\phi+\sum_{k}\log S^{21}(\phi-\thetak-\tfrac{\mathrm{i}\pi}{2})\nonumber \\
 & \qquad-\smash{\int\frac{d\theta}{2\pi}}\:K^{21}(\phi-\theta)L_{1}(\theta)\label{eq:tba-su32}
\end{align}
where $L_{i}(\theta)\equiv\log\left[1+e^{-\epsilon_{i}(\theta)}\right]$,
and the interaction kernel is nonzero only between particles of different
mass: 
\begin{align*}
K^{ij}(\theta) & \equiv-\mathrm{i}\partial_{\theta}\log S^{ij}(\theta).\\
 & =\frac{1}{\cosh(\theta\pm\sigma)},\quad i,j={\textstyle {1,2\atop 2,1}}.
\end{align*}
 The rapidities $\thetak$ of the physical particles are fixed in
terms of their mode numbers $\ennk\in\mathbb{Z}$ by

\begin{equation}
\epsilon_{1}(\thetak+\tfrac{\mathrm{i}\pi}{2})=2\pi\mathrm{i}(\ennk+\tfrac{1}{2}).\label{eq:quant-cond}
\end{equation}
The purpose of solving these integral equations is to find the energy
\begin{equation}
E=\sum_{k}m_{1}\cosh\thetak-\smash{\sum_{i=1,2}\int\frac{d\theta}{2\pi}}\:m_{i}\cosh\theta\:L_{i}(\theta).\label{eq:energy-su32}
\end{equation}

Dropping all the integrals will convert (\ref{eq:quant-cond}) into
the Bethe Ansatz Equations (BAE), momentum quantisation conditions
for otherwise free particles \citep{Bethe:1931hc,Faddeev:1977rm}.
Here these are simply $e^{\mathrm{i}m_{1}\sinh\thetak^{\mathrm{BAE}}}=\prod_{k'\neq k}(-1)$.
This approximation is usually justified when $L$ is large, as we
generically expect $\epsilon_{i}(\theta)$ to be large, and hence
the factor $L_{i}(\theta)$ which appears in every integral to be
small. For a massive theory, it gives the energy to polynomial accuracy,
i.e. including all terms in $1/L$.

The first corrections for smaller $L$ are the Lüscher terms \citep{Luscher:1983rk}.
In deriving these from the TBA there are two contributions \citep{Bajnok:2008bm}:
$\delta E_{\mathrm{int}}$ is the integrals in energy (\ref{eq:energy-su32}),
and 
\[
\delta E_{\mathrm{quant}}\equiv{\textstyle \sum_{k}}m_{1}\cosh\thetak-{\textstyle \sum_{k}}m_{1}\cosh\thetak^{\mathrm{BAE}}
\]
comes from the integral in (\ref{eq:tba-su32}) via the quantisation
condition (\ref{eq:quant-cond}). Both enter with a factor $e^{-m_{i}L}$,
which ensures that the wrapping effect of a massive particle is exponentially
suppressed; this may be thought of as a tunneling effect. Terms with
$e^{-2m_{i}L}$ are called double-wrapping effects.

\begin{figure}
\centering \includegraphics[width=1\columnwidth]{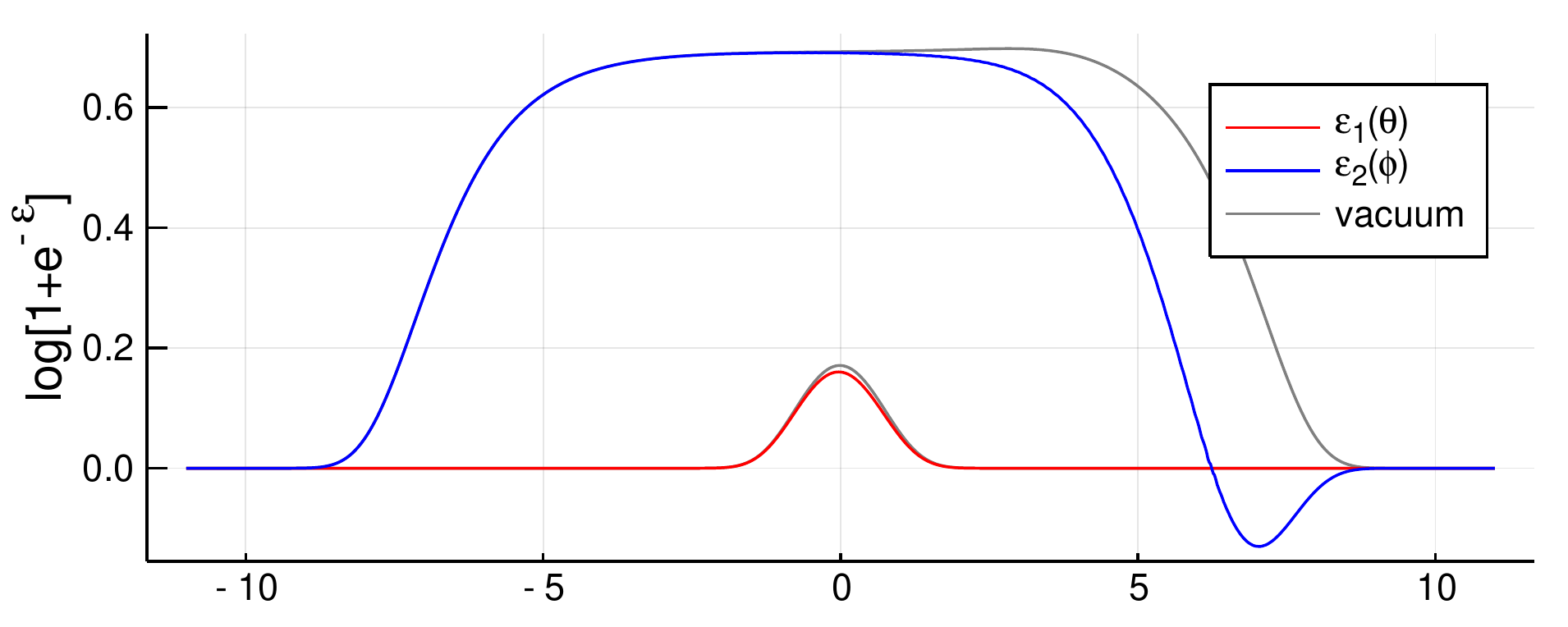}

\caption{%\footnotesize 
A solution of the $su(3)_{2}$ model (\ref{eq:tba-su32}), with $m_{1}=1$,
$m_{2}=10^{-3}$ and $\sigma=4$, in length $L=2$. The vacuum solution
is shown in gray, while the coloured lines are an excited state with
$\ennone=1$ hence $\thetaone=2.238$. We are interested in much larger
$L$, for which the red line $L_{1}(\theta)\sim e^{-m_{1}L}\to0$,
and even smaller $m_{2}$ such that $m_{2}L\to0$. Then the blue line
$L_{2}(\phi)$ is order 1 for $\left|\phi\right|\lesssim-\log m_{2}L\to\infty$.
\label{fig:epsilon-massive-massless}}
\end{figure}
However, Lüscher terms arising from massless virtual particles need
not be so suppressed. Their exponent contains $m_{2}L\cosh\phi$ which
approaches $L\left|p\right|$ in the limit $m_{2}\to0$, giving polynomial
corrections like $\int dp\:e^{-L\left|p\right|}=2/L$. It is the momentum
$p=m_{2}\sinh\phi$ which is well-defined in the massless limit, while
$\phi$ diverges. Figure \ref{fig:epsilon-massive-massless} shows
a comparison between massive $L_{1}(\theta)$ and near-massless $L_{2}(\phi)$.

In many theories this divergence of rapidity $\phi$ would cause the
interaction $S^{12}$ to become trivial. But in HSG models it is possible
to compensate with the shift parameter $\sigma$, and so we propose
taking this double-scaling limit:
\begin{equation}
m_{2}\to0,\quad\sigma\to+\infty,\quad\sigma_{0}=\sigma+\log(m_{2}/2)\sim1.\label{eq:m2-zero-limit}
\end{equation}
Our choice of the sign of $\sigma$ means that it is the right-moving
particles which retain an interaction:
\[
S^{21}(\phi-\thetak-\tfrac{\mathrm{i}\pi}{2})\to\begin{cases}
-1\big/\tanh\frac{\log p-\thetak-\sigma_{0}}{2}, & p>0\\
+1 & p<0.
\end{cases}
\]
Dropping all order $e^{-m_{1}L}$ pieces, the integral in (\ref{eq:energy-su32})
arising from massless virtual particles then reads 
\begin{align}
\negthickspace\delta E_{\mathrm{int}}\negthickspace & =\frac{-\pi}{24L}-\negthickspace\int_{0}^{\infty}\negthickspace\frac{dp}{2\pi}\log\Big[1-e^{-L|p|}\prod_{k}\tanh\tfrac{\log p-\thetak-\sigma_{0}}{2}\Big]\nonumber \\
 & =\frac{\pi}{12}\Big[-\frac{1}{L}+\frac{1}{L^{2}}-\frac{2+c_{1}}{L^{3}}+\mathcal{O}\Big(\frac{1}{L^{4}}\Big)\Big]\label{eq:luscher-simple-model}
\end{align}
at $\sigma_{0}=0$, for one particle at $\thetaone=c_{1}/L+c_{2}/L^{2}+\ldots$.
This expansion can be checked against a numerical solution of the
full TBA (\ref{eq:tba-su32}) at small but finite $m_{2}=10^{-6}$,
and we see perfect agreement. Analytically, notice that if we expand
in the wrapping number (i.e. in $e^{-L\left|p\right|}$) then every
wrapping contributes at order $1/L$. However, expanding the integrand
in $1/L$, holding fixed $y=pL$, gives the series shown. 

The quantisation condition (\ref{eq:quant-cond}) can be treated in
the same way. The effect of massless virtual particles enters $\thetak$
at order $1/L^{2}$, and hence affects the energy as $\delta E_{\mathrm{quant}}\propto c_{2}/L^{3}$.

\subsection*{$\mathbf{su(3)_{3}}$ Model}

\noindent What the above $su(3)_{2}$ example lacks is interactions
between the massless particles. We next turn to the $su(3)_{3}$ HSG
model, which has the same three parameters $m_{1}$, $m_{2}$, $\sigma$
but now two particles of each mass, $a,b=1,2$. (It also has a discrete
parameter, a 3rd root of $-1$, which we take to be $\eta=-1$ for
simplicity.) The S-matrix is again diagonal, and we write $S_{ab}^{ij}(\theta_{a,i}-\theta_{b,j})$
for a particle of mass $m_{i}$ and label $a$ scattering with one
of $m_{j},b$. This is
\begin{align*}
S^{ij}(\theta)\,= & \delta_{ij}\,\begin{bmatrix}(2)_{\theta} & -(1)_{\theta}\\
-(1)_{\theta} & (2)_{\theta}
\end{bmatrix}\\
 & +\bigl(1-\delta_{ij}\bigr)\,\begin{bmatrix}-(-1)_{\theta+\sigma_{ij}} & (-2)_{\theta+\sigma_{ij}}\\
(-2)_{\theta+\sigma_{ij}} & -(-1)_{\theta+\sigma_{ij}}
\end{bmatrix}
\end{align*}
where $\sigma_{12}=-\sigma_{21}=\sigma$ and 
\[
(n)_{\theta}\equiv\sinh\tfrac{1}{2}(\theta+\tfrac{\mathrm{i}\pi}{3}n)/\sinh\tfrac{1}{2}(\theta-\tfrac{\mathrm{i}\pi}{3}n).
\]
 The complete TBA has four pseudo-energies $\epsilon_{a,i}(\theta)$,
and we again consider an excited-state TBA with physical particles
of mass $m_{1}$, and label $a=1$. This reads
\begin{align}
\epsilon_{a,i}(\theta)\,= & m_{i}L\cosh\theta+\sum_{k}\log S_{a1}^{i1}(\theta-\thetak-\tfrac{\mathrm{i}\pi}{2})\nonumber \\
 & -\smash{\sum_{b,j}\int\frac{d\theta'}{2\pi}}\;K_{ab}^{ij}(\theta-\theta')L_{b,j}(\theta')\label{eq:pseudo-en-original}
\end{align}
 with energy
\begin{align}
E & =\sum_{k}m_{1}\cosh\thetak-\sum_{b,j}\int\frac{d\theta}{2\pi}\:m_{j}\cosh\theta\:L_{b,j}(\theta).\label{eq:energy-su33}
\end{align}
The quantisation condition for $\thetak$ is now $\epsilon_{1,1}(\thetak+\tfrac{\mathrm{i}\pi}{2})=2\pi\mathrm{i}(\ennk+\tfrac{1}{2})$,
which in the large-$L$ limit gives us the following Bethe equations:
\[
e^{\mathrm{i}m_{1}L\sinh\thetaj^{\mathrm{BAE}}}{\textstyle \prod_{k}}\:S_{11}^{11}(\thetaj^{\mathrm{BAE}},\thetak^{\mathrm{BAE}})=-1.
\]

\begin{figure}
\includegraphics[width=1\columnwidth]{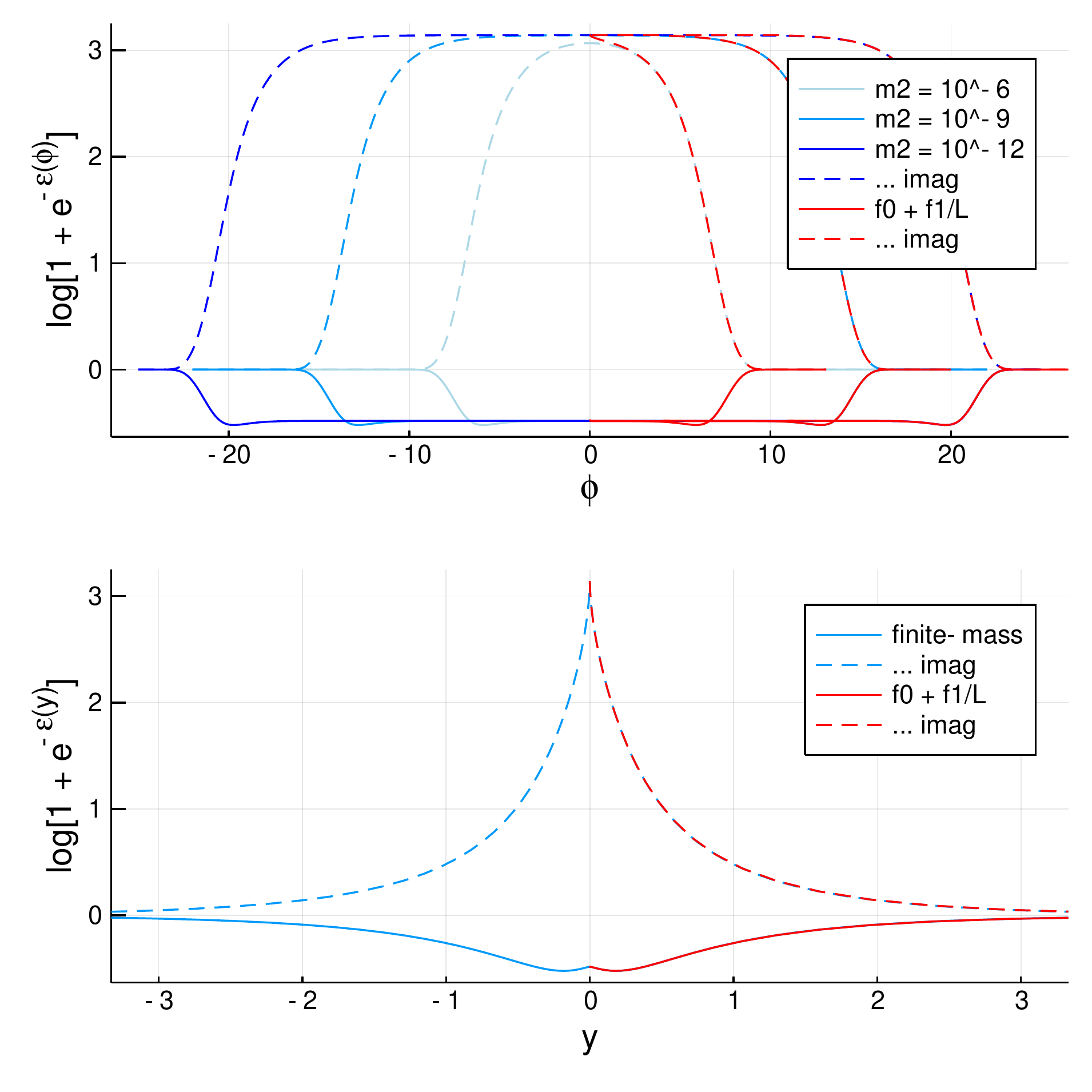}

\caption{%\footnotesize 
 Pseudo-energy $\epsilon_{1,2}(\phi)$ for the $su(3)_{3}$ HSG model,
showing the solution to (\ref{eq:pseudo-en-with-m2}) in length $L=10^{3}$
with masses $m_{2}=10^{-6},10^{-9},10^{-12}$ (in blue), and on the
same axes as the zero-mass $f_{n}(y)$ expansion (\ref{eq:f-first-eqs})
(drawn for $y>0$ only, in red). As a function of $\phi$ (above)
the range for which $L_{1,2}(\phi)$ is order 1 grows as the mass
$m_{2}$ is decreased. As a function of $y=m_{2}L\sinh\phi$ (below)
the massless limit is well-behaved. We use mode number $\ennone=1$
hence $\thetaone=4\pi/L+\ldots$, mass $m_{1}=1$, shift $\sigma=-\log m_{2}/2$.
Solid lines are the real parts, dashed lines imaginary.\label{fig:epsilon(phi)-and-(y)}}
\end{figure}

\subsection*{Massless TBA}

\noindent To study the TBA in the $m_{2}\to0$ limit, we now fix
$m_{1}\approx1$ and drop all exponentially small terms, that is,
all integrals containing $L_{a,1}(\theta)$ above. Because we have
only physical particles of type $a=i=1$, the two massless equations
are complex conjugates, $\epsilon_{2,2}(\phi)=\overline{\epsilon_{1,2}}(\phi)$,
leaving just one integral equation:
\begin{align}
 & \epsilon_{1,2}(\phi)=m_{2}L\cosh\phi-{\textstyle \sum_{k}}\log s(\phi-\thetak-\sigma)+\mathrm{i}\pi\label{eq:pseudo-en-with-m2}\\
 & +\sqrt{3}\int\frac{d\phi'}{2\pi}\left[\frac{L_{1,2}(\phi')}{1+2\cosh(\phi-\phi')}-\frac{\overline{L_{1,2}}(\phi')}{1-2\cosh(\phi-\phi')}\right]\nonumber 
\end{align}
where $s(\theta)\equiv(1)_{\theta-\frac{\mathrm{i}\pi}{2}}=\sinh(\tfrac{\theta}{2}-\tfrac{\mathrm{i}\pi}{12})/\sinh(\tfrac{\theta}{2}-\tfrac{\mathrm{i}5\pi}{12})$,
and 
\[
E=\sum_{k}m_{1}\cosh\thetak-\int\frac{d\phi}{2\pi}m_{2}\cosh\phi\left[L_{1,2}(\phi)+\overline{L_{1,2}}(\phi)\right].
\]
We can solve this numerically at small finite $m_{2}$, but can also
take a strict $m_{2}\to0$ limit analytically, using the same double-scaling
limit as above, (\ref{eq:m2-zero-limit}). It is again the mixed-mass
S-matrix elements which contain the shift $\sigma$, and choosing
to take $\sigma\to+\infty$ keeps the coupling to right-moving massless
modes nontrivial:
\[
s(\phi-\thetak-\sigma)\to s_{+}(p,\thetak)\equiv\begin{cases}
s(\log p-\thetak-\sigma_{0}), & p>0\\
e^{-\mathrm{i}\pi/3} & p<0.
\end{cases}
\]
Notice that in this limit, where $\left|\phi\right|,\left|\phi'\right|\to\infty,$
the denominators of (\ref{eq:pseudo-en-with-m2}) diverge unless $\phi$
and $\phi'$ have the same sign. Hence the integral equation for $\epsilon(p),p>0$
is decoupled from that for $p<0$:
\begin{align}
 & \epsilon_{1,2}(p)=L\left|p\right|-{\textstyle \sum_{k}}\log s_{+}(p,\thetak)+\mathrm{i}\pi\label{eq:pseud-en-in-p}\\
 & +\frac{\sqrt{3}}{2\pi}\int_{0}^{\infty}\frac{dp'}{p'}\left[\frac{L_{1,2}(\pm p')}{1+\frac{\left|p\right|}{p'}+\frac{p'}{\left|p\right|}}-\frac{\overline{L_{1,2}}(\pm p')}{1-\frac{\left|p\right|}{p'}-\frac{p'}{\left|p\right|}}\right],\;p\gtrless0.\nonumber 
\end{align}
The energy contains $\delta E_{\mathrm{int}}=-\int\frac{dp}{2\pi}\left[L_{1,2}(p)+\mathrm{c.c.}\right]$,
notice the different measure. And the quantisation condition (for
one physical particle, $\thetaone\in\mathbb{R}$) reads
\begin{align}
2\pi\mathrm{i}\ennone & =\mathrm{i}m_{1}L\sinh\thetaone-2\pi\mathrm{i}\label{eq:quant-cond-in-p}\\
 & +\frac{\sqrt{3}}{2\pi}\int_{0}^{\infty}\frac{dp}{p}\left[\frac{\log(1+e^{-\epsilon_{1,2}(p)})}{1-2\mathrm{i}\sinh(\thetaone-\log p+\sigma_{0})}-\mathrm{c.c.}\right].\nonumber 
\end{align}

\begin{figure}
\centering \includegraphics[width=1\columnwidth]{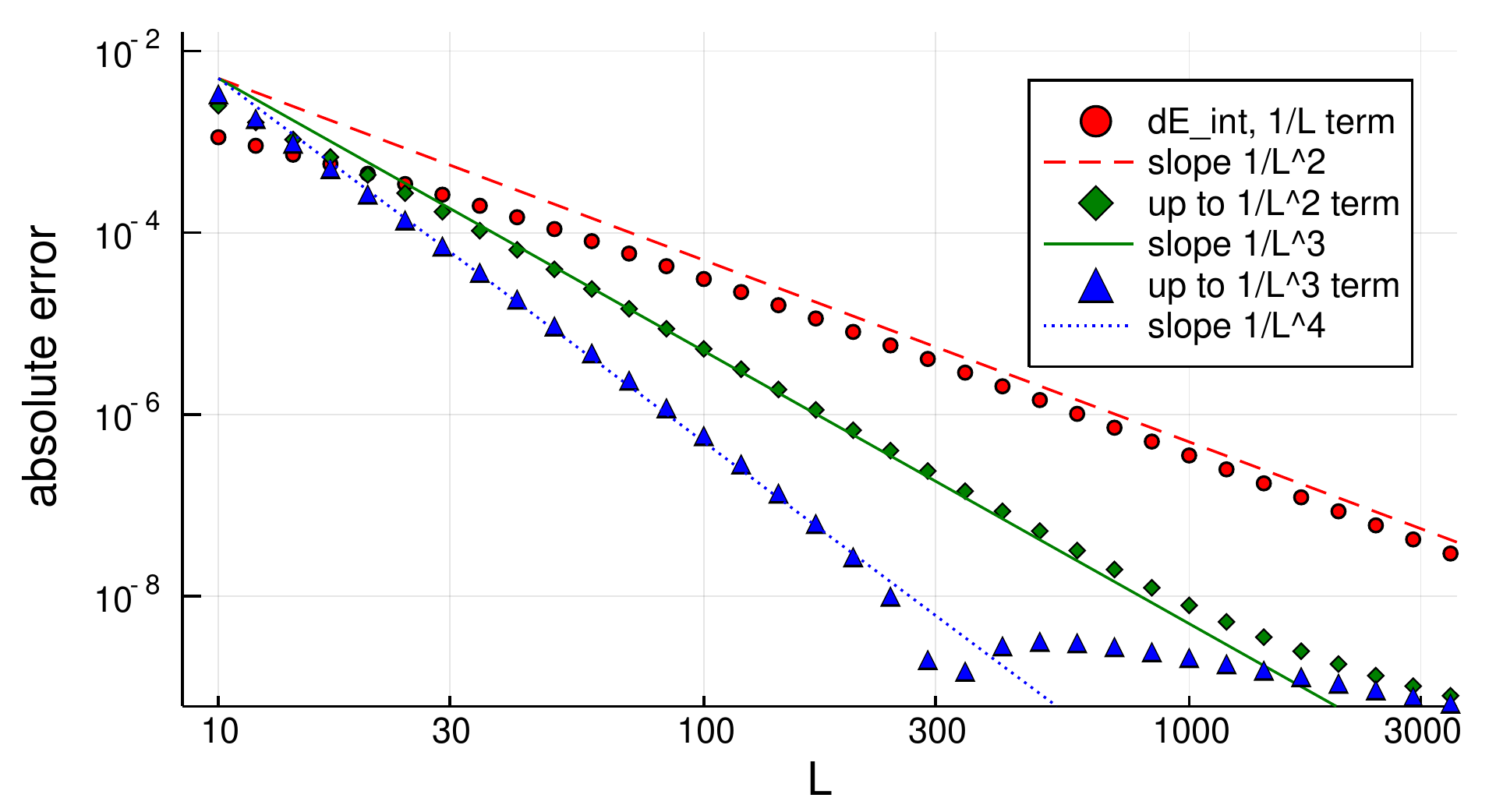}

\caption{%\footnotesize 
Comparison of the energy correction $\delta E_{\mathrm{int}}$ from
the $m_{2}=0$ expansion (\ref{eq:f-first-energies}), to that from
solving the TBA (\ref{eq:pseudo-en-with-m2}) at $m_{2}=10^{-12}$.
Including more terms (up to $1/L^{3}$) reduces the error, at moderately
large $L$. (At around $L=300$ the error changes sign, and we run
out of precision.) \label{fig:energy-errors}}
\end{figure}

This limit eliminates $m_{2}$ but not $L$ from the integral equation.
To find large-$L$ solutions, we claim that you should expand in $1/L$
holding fixed $pL$, the same small-momentum limit we used for (\ref{eq:luscher-simple-model})
above:

\begin{align}
\epsilon_{1,2}(p) & =f_{0}(y)+\frac{f_{1}(y)}{L}+\frac{f_{2}(y)}{L^{2}}+\mathcal{O}\Big(\frac{1}{L^{3}}\Big)\label{eq:f-ansatz}\\
\thetaone & =c_{1}/L+c_{2}/L^{2}+\ldots,\qquad y=pL.\nonumber 
\end{align}
This ansatz gives an integral equation at each power of $1/L$. For
$y<0$, clearly (\ref{eq:pseud-en-in-p}) is independent of $L$,
hence only $f_{0}(y)$ is nonzero there. The first few equations are:\begin{widetext}
\begin{align}
f_{0}(y) & =\left|y\right|+2\pi\Big(1-\frac{\mathop{\mathrm{sign}}(y)}{3}\Big)+\frac{\sqrt{3}}{2\pi}\int_{0}^{\infty}\frac{dy'}{y'}\Bigg[\frac{\log(1+e^{-f_{0}(\pm y')})}{1+\frac{\left|y\right|}{y'}+\frac{y'}{\left|y\right|}}-\frac{\log(1+e^{-\overline{f_{0}}(\pm y')})}{1-\frac{\left|y\right|}{y'}-\frac{y'}{\left|y\right|}}\Bigg],\negthickspace\negthickspace\negthickspace & \negthickspace\negthickspace\negthickspace y & \gtrless0\label{eq:f-first-eqs}\\
f_{1}(y) & =\frac{\sqrt{3}y}{2}-\frac{\sqrt{3}}{\pi}\int_{0}^{\infty}\frac{dy'}{y'}\Big[\frac{f_{1}(y')}{(1+e^{+f_{0}(y')})(1+\frac{y}{y'}+\frac{y'}{y})}-\frac{\bar{f_{1}}(y')}{(1+e^{+\bar{f_{0}}(y')})(1-\frac{y}{y'}-\frac{y'}{y})}\Big], & \negthickspace\negthickspace\negthickspace y & >0\nonumber \\
f_{2}(y) & =\frac{-\mathrm{i}\sqrt{3}y^{2}}{2}-\sqrt{3}yc_{1}+\frac{\sqrt{3}}{4\pi}\int_{0}^{\infty}\frac{dy'}{y'}\Big[\frac{e^{+f_{0}(y')}f_{1}(y')-2(1+e^{+f_{0}(y')})f_{2}(y')}{(1+e^{+f_{0}(y')})^{2}(1+\frac{y}{y'}+\frac{y'}{y})}-...\Big], & \negthickspace\negthickspace\negthickspace y & >0.\nonumber 
\end{align}
\end{widetext}These can be solved in sequence, as each depends only
on lower-order functions $f_{n}(y)$, and lower-order coefficients
$c_{n}$. The resulting pseudo-energy $\epsilon_{1,2}(\phi)$ is shown
in Figure \ref{fig:epsilon(phi)-and-(y)}. On the same axes we show
the result of solving (\ref{eq:pseudo-en-with-m2}) at small but finite
$m_{2}$; see the appendix for a discussion of numerical issues here.
The coefficients of $\thetaone$ are found by expanding (\ref{eq:quant-cond-in-p}),
and solving: 
\begin{align}
\negthickspace c_{1} & =\frac{2\pi}{m_{1}}(\ennone+1) & \negthickspace\negthickspace\negthickspace & =4\pi\label{eq:c-first-few}\\
\negthickspace c_{2} & =-\frac{\sqrt{3}}{m_{1}}\smash{\int}\frac{dy}{2\pi}\left[\log(1+e^{-f_{0}(y)})+\text{c.c.}\right] & \negthickspace\negthickspace\negthickspace & \approx0.362\nonumber \\
 &  & \negthickspace\negthickspace\negthickspace\negthickspace\negthickspace\negthickspace c_{3} & \approx-336.6\,.\nonumber 
\end{align}
The same numerical solutions $f_{n}(y)$ give the values shown (for
mode number $\ennone=1$, mass $m_{1}=1$, shift $\sigma_{0}=0$),
and again these agree with the TBA at small finite $m_{2}$. These
corrections to the quantisation condition for $\thetaone$ are (through
the $m_{1}\cos\thetaone$ term) one source of corrections to the energy:
\begin{equation}
\delta E_{\mathrm{quant}}=\frac{c_{1}c_{2}}{L^{3}}+\ldots\approx\frac{4.558}{L^{3}}+\mathcal{O}\Big(\frac{1}{L^{4}}\Big).\label{eq:dE_quant}
\end{equation}
The other source is the integral term in (\ref{eq:energy-su33}),
which can be similarly expanded: 
\begin{align}
\delta E_{\mathrm{int}} & =-\smash{\frac{1}{L}\int\frac{dy}{2\pi}}\left[\log(1+e^{-f_{0}(y)})+\text{c.c.}\right]\nonumber \\
 & \qquad\qquad+\frac{1}{L^{2}}\int\frac{dy}{2\pi}\left[\frac{f_{1}(y)}{1+e^{f_{0}(y)}}+\text{c.c.}\right]+\ldots\nonumber \\
 & \approx\frac{0.418}{L}-\frac{0.363}{L^{2}}+\frac{5.855}{L^{3}}+\mathcal{O}\Big(\frac{1}{L^{4}}\Big).\label{eq:f-first-energies}
\end{align}
We are able to check the first three terms here against the finite-$m_{2}$
numerical solution to (\ref{eq:pseudo-en-with-m2}), and Figure \ref{fig:energy-errors}
shows the comparison. 

For contrast, we can also find the Lüscher contribution here (instead
of solving the integral equation) by dropping the integral in (\ref{eq:pseud-en-in-p}).
The answer is very different:
\begin{align}
\delta E_{\mathrm{int}} & =-\int\frac{dp}{2\pi}\left[\log\big(1-e^{-\left|L\right|p}s_{+}(p,\thetaone)\big)+\mathrm{c.c.}\right]\nonumber \\
 & \approx-\frac{0.349}{L}+\frac{0.302}{L^{2}}+\mathcal{O}\Big(\frac{1}{L^{3}}\Big).\label{eq:su33-luscher}
\end{align}

\subsection*{Conclusion}

\noindent By studying these simple models, we have learned:
\begin{enumerate}
\item Interacting relativistic integrable theories containing both massive
and massless particles exist \footnote{We exclude here what are variously called pseudoparticles \citep{Fendley:1992dm},
auxiliary roots \citep{Beisert:2005fw}, or magnon Bethe roots \citep{Gromov:2008gj}.
The massless excitations of interest are ordinary, propagating, momentum-carrying
particles, on an equal footing with the massive ones.}.
\item The spectrum of massive excitations depends on the massless sector,
including massless-massless interactions. Calculating this $E$ to
polynomial accuracy in $L$, which in a massive theory requires only
the BAE, now requires at least the massless-sector TBA.
\item Either left- or right-moving massless particles could be nontrivially
coupled to the massive modes, but not both. We expect that more complicated
theories can somewhat avoid this \footnote{For example the $su(4)_{k}$ HSG models have three mass parameters
$m_{i}$ for $i=1,2,3$, and three independent shifts $\sigma_{ij}$
for $i>j$, allowing (say) $m_{3}=0$ particles moving in either direction
to remain coupled, but not coupled to the same massive species.}. 
\item While an expansion in the wrapping order (i.e. in $e^{-L\left|p\right|}$)
is no longer meaningful, the energy calculation can be organised as
a series in $1/L$ by expanding at small momentum, holding $pL$ fixed.
\end{enumerate}
As mentioned above, our motivation for this work comes from string
integrability in the AdS/CFT correspondence. The light-cone gauge
string is viewed as a non-relativistic integrable field theory, living
on the worldsheet whose spatial extent $L$ is proportional to an
angular momentum $J$. And in $AdS_{3}\times S^{3}\times T^{4}$ this
theory has massless excitations.

In earlier work \citep{Abbott:2015pps}, we showed that some disagreements
in the one-loop spectrum of the massive sector $AdS_{3}\times S^{3}$
appear to be caused by interactions with the massless sector. In particular,
we were able to calculate massless Lüscher corrections for circular
spinning strings, for which there was a long-standing mismatch. We
included all orders of wrappings following \citep{Heller:2008at},
and treated the multi-particle physical state following \citep{Bajnok:2008bm},
to calculate \footnote{Here $b$ labels the 4 massless bosons and 4 massless fermions, and
$n$ is the wrapping order. The physical solution is a condensate
of a very large number, of order $\sqrt{\lambda}$, of massive particles,
all of the same type, which can be treated as a single cut in the
complex plane \citep{Hernandez:2006tk}. The 't Hooft coupling is
$\lambda=R^{4}/\alpha^{\prime2}$ in terms of the radius of the $AdS$
space, and $1/\sqrt{\lambda}$ plays the role of $\hbar$ for these
semiclassical strings. }:
\[
\delta E=\sum_{b}^{4+4}(-1)^{F_{b}}\negthickspace\int\negthickspace d\phi\sum_{n=1}^{\infty}\frac{1}{n}e^{-nLm_{2}\cosh\phi}\prod_{k}^{\negthickspace\negthickspace\mathcal{O}(\sqrt{\lambda})\negthickspace\negthickspace}S_{b1}(\phi-\thetak)^{n}.
\]
This formula allowed us to correct the mismatch between BAE and string
theory calculations (up to a factor of 2) for strings moving in $S^{3}$,
called the $su(2)$ sector. However, there are comparable mismatches
for other solutions, such as $sl(2)$-sector circular strings, for
which a similar calculation does not succeed. This formula is the
analogue of (\ref{eq:luscher-simple-model}) or (\ref{eq:su33-luscher})
here. What we add now is the first glimpse of the world beyond these
wrapping corrections: the interactions of massless modes with each
other lead to different results.

Since our paper \citep{Abbott:2015pps}, there has been some work
on the massless TBA \citep{Bombardelli:2018jkj,Fontanella:2019ury}.
Unlike the massive sector, there appear to be no complications with
massless bound states. These papers take a small-$p$ limit in which
the system becomes relativistic. However they do not yet incorporate
massive-massless interactions, which are essential for the effects
on the massive spectrum studied here. It would be very interesting
to find ways to remedy this.

\subsection*{Acknowledgements}

\noindent We thank Zoltan Bajnok, Patrick Dorey, Romuald Janik, and
Luis Miramontes for helpful conversations.

MCA is supported by a Wigner Fellowship, and NKIH grant FK 128789.
IA is supported by an EPSRC Early Career Fellowship EP/S004076/1.
At an earlier stage of this work, both were supported by Polish NCN
grant 2012/06/A/ST2/00396.

\bibliographystyle{my-JHEP-4}
\bibliography{/Users/me/Dropbox/Library.papers3/complete3}

\appendix
% \vspace{1cm}
\newpage 

\section*{Numerical Methods}

\noindent We solve all of these integral equations iteratively, starting
with $\epsilon(\phi)=0$ and then at time step $t$ replacing 
\[
\epsilon(\phi):=\lambda^{t}\mathop{\mathrm{rhs}}[\epsilon](\phi)+(1-\lambda^{t})\epsilon(\phi).
\]
Here $\lambda=0.99$ controls how fast the updates decay. Some such
damping is essential in order to find stable and accurate solutions.
The function $\epsilon(\phi)$ is encoded either as a sum of Chebyshev
polynomials, or just values at a grid of points $\phi_{i}$.

\begin{figure}
\includegraphics[width=1\columnwidth]{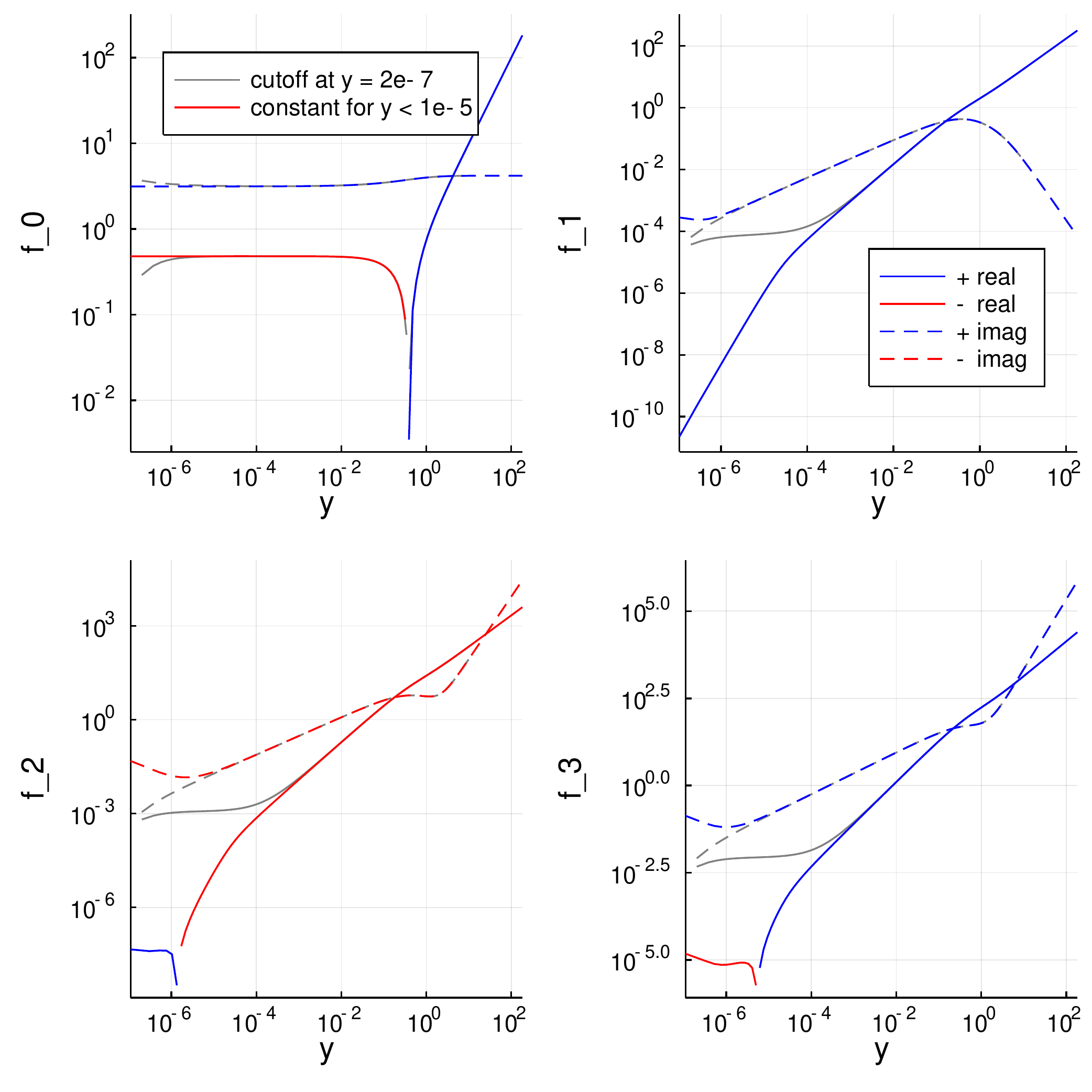}

\caption{%\footnotesize 
 Terms $f_{n}(y)$ of the expansion (\ref{eq:f-ansatz}). The lines
in colour use $f_{0}(y)=-0.48+\mathrm{i}\pi$ for $y<10^{-5}$, while
the lines in gray do not, and hence suffer more strongly from boundary
effects. Dashed lines indicate the imaginary part, and red indicates
negative values. \label{fig:f_n(y)}}
\end{figure}
The equations (\ref{eq:f-first-eqs}) for $f_{n}(y)$ deserve extra
comment. We use a grid of points evenly spaced in $\phi$, which necessarily
has some smallest value $y_{\mathrm{min}}$. The integration measure
$\int dy'/y'$ is uniform in $\log y$ and thus arbitrarily small
$y'$ are weighted equally. In particular, the values $y'<y_{\mathrm{min}}$
which we omit would appear inside the integral needed for $f_{n}(y>y_{\mathrm{min}})$.
This effect is of finite range in $\log y$, thanks to factors $1/[1\pm2\cosh(\log y-\log y')]$
in the integrand, and thus we believe it contaminates a only the end
of the range of values considered.

Figure \ref{fig:f_n(y)} shows these boundary effects. With a cutoff
$y_{\mathrm{min}}=2\times10^{-7}$, values of $f_{0}(y)$ for $y<10^{-6}$
deviate, but it is simple to correct this by solving the small-$y$
limit directly:
\begin{align*}
f_{0}(0) & =\frac{4\mathrm{i}\pi}{3}+I_{+}\log(1+e^{-f_{0}(0)})-I_{-}\log(1+e^{-\bar{f_{0}}(0)})\\
 & \approx-0.48+\mathrm{i}\pi
\end{align*}
where $I_{\pm}\equiv\frac{\sqrt{3}}{2\pi}\int_{0}^{\infty}\frac{dy}{y}\frac{1}{1\pm(y+1/y)}$
with no cutoff. Fixing $f_{0}(y)$ to be exactly this for $y<10^{-5}$
obviously produces a perfectly straight line in both Figure \ref{fig:f_n(y)}
and Figure \ref{fig:epsilon(phi)-and-(y)}. The other $f_{n}(y)$
have similar boundary effects which are not so easily removed, but
the effect of clamping $f_{0}(y)$ to a constant is the difference
between the gray and coloured curves in Figure \ref{fig:f_n(y)}.

The energy integrals (\ref{eq:f-first-energies}), which are the motivation
for finding functions $f_{n}(y)$, have a different integration measure,
simply $\int dy$. Thus very small $y$ values contribute vanishingly
little to $\delta E_{2}$. The digits shown for $\delta E_{2}$ in
(\ref{eq:f-first-energies}) above are unchanged by clamping $f_{0}(y)$
to a constant like this, and also unchanged by only integrating over
$y>10^{-3}$.

The numerical solution also needs a cutoff $y_{\mathrm{max}}$. Here
there are no awkward issues, as the integrand always has a factor
$e^{-f_{0}(y)}$ which rapidly kills it. This is true for both $f_{n}(y)$
integral equations (\ref{eq:f-first-eqs}), and for the energy integrals
(\ref{eq:f-first-energies}).

\begin{figure}
\centering \includegraphics[width=1\columnwidth]{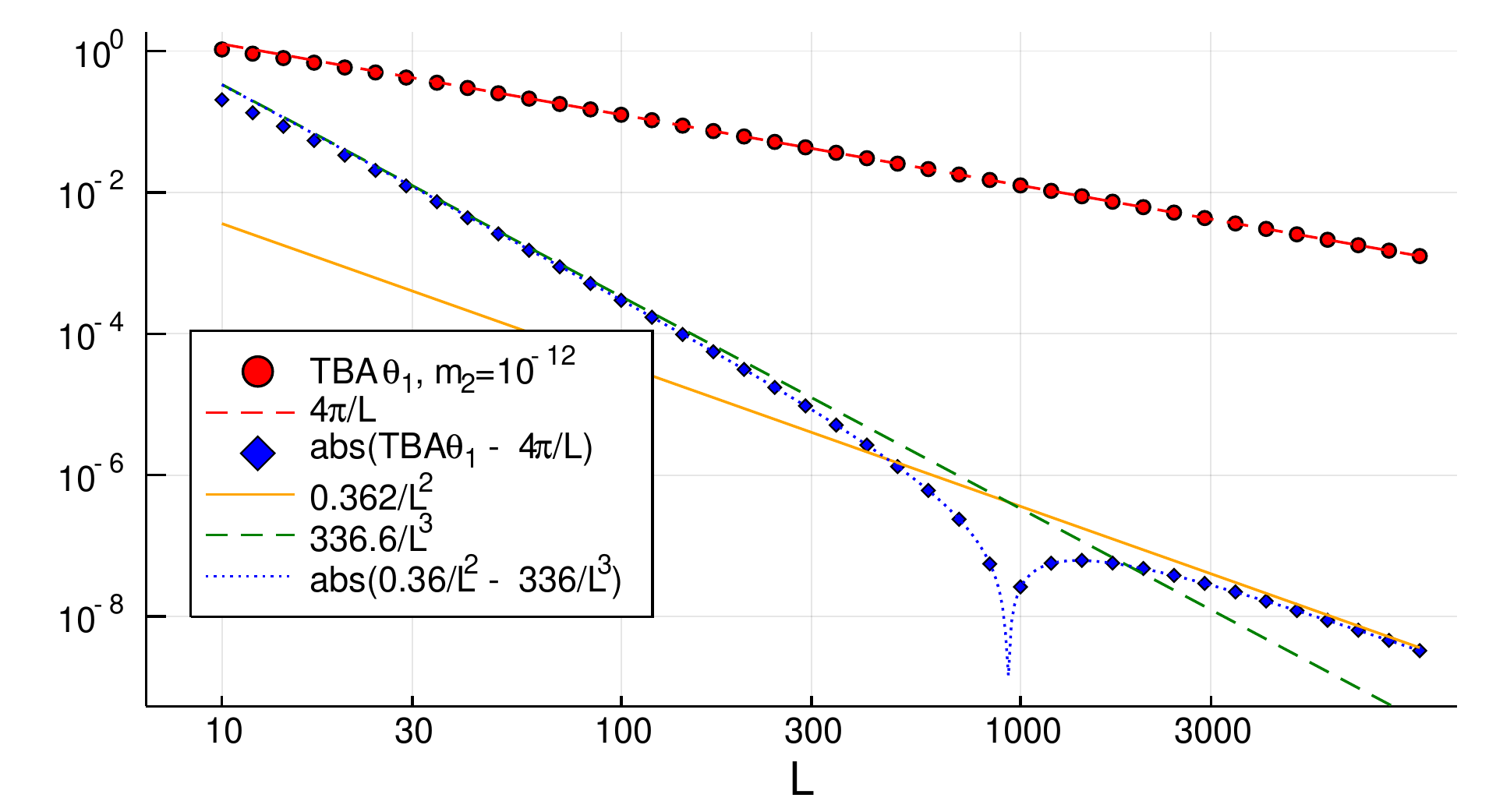}

\caption{%\footnotesize 
Comparison of $\thetaone$ from solving (\ref{eq:pseudo-en-with-m2})
at $m_{2}=10^{-12}$, to the expansion $c_{1}/L+c_{2}/L^{2}+c_{3}/L^{3}+\ldots$,
using the first three $c_{n}$ as shown in (\ref{eq:c-first-few}).
The difference $\thetaone-4\pi/L$ changes sign around $L=1000$,
causing a dip in the blue points. \label{fig:numerical-theta-cn-comparison}}
\end{figure}

Finally, in the text we claim agreement between the coefficients $c_{n}$
found along with these $f_{n}(y)$, shown in (\ref{eq:c-first-few}),
and the result of solving the TBA (\ref{eq:pseudo-en-with-m2}) at
small but finite $m_{2}$. Figure \ref{fig:numerical-theta-cn-comparison}
shows some data on this.

\end{document}